\documentclass[prb, fleqn,twocolumn, showpacs, showkeys]{revtex4}% Physical Review B
\usepackage[dvips]{epsfig}

\newcommand{\Spin}[1]{\mathbf{S_{#1}}}
\newcommand{\vectL}[1]{\mathbf{L_{#1}}}
\newcommand{\vectLnorm}{\mathbf{l}}
\newcommand{\vectLnormI}[1]{\mathbf{l_{#1}}}
\begin{document}
\title{Magnetic phase diagram and structure of the magnetic phases in the quasi-one-dimensional antiferromagnet
 BaCu$_2$Si$_2$O$_7$: symmetry analysis.}
\author{V. Glazkov}
    \affiliation{P.L.Kapitza Institute for Physical Problems,
    117334 Moscow, Russia}
    \email{glazkov@kapitza.ras.ru}
\author{H.-A. Krug von Nidda}
    \affiliation{Experimentalphysik V, EKM, Institut f\"ur Physik, Universit\"at
    Augsburg, 86135 Augsburg, Germany}
\begin{abstract}
We have performed a symmetry analysis of the properties of the
recently discovered quasi-one-dimensional compound
BaCu$_2$Si$_2$O$_7$. The existence of the unusual
spin-reorientation transitions is explained as an effect of the
unusually strong relativistic interactions. The possible
connection between the magnitude of the relativistic interactions
and the low-dimensional structure of the BaCu$_2$Si$_2$O$_7$ is
discussed. The structure of the magnetic phases is determined.
\end{abstract}

\date{\today}

%75.30.Kz Magnetic phase boundaries (including magnetic
%transitions, metamagnetism, etc.)
%75.50.Ee Antiferromagnetics



\maketitle

\section{\label{sec:intro}Introduction.}

Low-dimensional magnets are subject of intense interest due to the
important role of the quantum fluctuations in these systems. As it
was rigorously shown\cite{mermin-wagner}, no antiferromagnetic
order can exist in an one-dimensional system. However, a
weak interchain interaction leads to the formation of long-range
magnetic order.

Different types of magnetic order (differing by the
relative orientation of the spins) are possible if the
unit cell of an antiferromagnet contains more than two
magnetic ions.  A transition from the paramagnetic phase to the phase with
lowest exchange energy occurs at the N\'{e}el
point. The difference between energies of different types of magnetic order
in usual 3D-antiferromagnets
is of the scale of the exchange interaction $J$.

Quasi-one-dimensional antiferromagnets have some unique
properties: First, if the order types differ only by the
orientations of the spins in the directions perpendicular to the
chain, the energy difference between these ordered states will be
small (of the scale of the small interchain-exchange interaction
$J_{\perp}$). A second peculiarity arises from the microscopic
structure of low-dimensional magnets. Small values of the
interchain-exchange constants in inorganic compounds are
frequently due to the orthogonality of the corresponding electron
orbitals. Since this "90$^{\circ}$"-rule is not applicable to
other interchain interactions, relativistic interchain
interactions (such as anisotropic or Dzyaloshinskii-Moriya
exchange interactions) can compete with the interchain
Heisenberg-exchange interaction. These features of low-dimensional
magnets can result in magnetic phase transitions with a change of
the exchange structure, as in the case of La$_2$CuO$_4$.
\cite{borovikkreines}

The recently discovered compound BaCu$_2$Si$_2$O$_7$ has attracted
attention, since it was reported that it is a
quasi-one-dimensional antiferromagnet \cite{tsukada:si}. The
CuO$_4$ plaquettes form zigzag chains along the $c$ direction of
the orthorhombic crystal with a Cu-O-Cu bond angle of about
124$^{\circ}$. The exchange constants determined by means of
neutron scattering\cite{kenzelmann} are $J_c=24.1$~meV,
$J_b=0.20$~meV, $J_a=-0.46$~meV and $J_{[110]}=0.15$~meV. The
susceptibility \cite{tsukada:si} $\chi(T)$ also demonstrates a
broad maximum characteristic for one-dimensional
antiferromagnets\cite{BonnerFisher} at a temperature near 200~K.

The weak interchain interaction leads to the formation of
long-range antiferromagnetic order at $T_N=9.2$~K. The N\'{e}el
temperature is well marked by a kink in $\chi_c$, indicating
easy-axis ordering in low magnetic field. Above $T_N$ the
susceptibility demonstrates a deviation from the Bonner-Fisher
law\cite{tsukada:si} --- $\chi_c$ strongly increases on
approaching the N\'{e}el temperature. The magnetic moment per
Cu-ion is strongly reduced in the ordered phase due to the effect
of quantum fluctuations: according to the inelastic neutron
scattering experiments\cite{zheludev:struct}
$\langle\mu\rangle=0.15\mu_B$.

Unexpectedly, the magnetization study in the ordered state reveals
the existence of two spin-reorientation transitions at $\mathbf{H}\parallel
c$ (transition fields $H_{c1}=2.0$~T and
$H_{c2}=4.9$~T).\cite{tsukada:2sf} Two spin-reorientation
transitions are a surprising feature for a
supposed-to-be-collinear antiferromagnet. Usually an easy-axis
antiferromagnet exhibits only one spin-reorientation transition
(spin-flop transition) caused by the competition of the gain in
magnetization energy with the loss in anisotropy energy. The
observation of an additional spin-reorientation
transition\cite{poirier:sf3} at $H\perp c$ at the field
$H_{c3}=7.7$~T is also intriguing --- the 'classical' easy-axis
antiferromagnet does not undergo a spin-reorientation transition,
if the external field is applied perpendicular to the easy axis of
the anisotropy. Neutron-scattering
experiments\cite{zheludev:struct} have shown that the
intermediate-field phase is noncollinear. However, the origin of
these exotic spin-reorientation transitions remains a mystery.

In the present paper we perform an analysis of the phase diagram
and  magnetic structure of BaCu$_2$Si$_2$O$_7$ from a macroscopic
approach taking into account the symmetry of the crystal. We show
that both two spin-reorientation transitions and the behavior of
the susceptibility above $T_N$ are due to the unusually large
relativistic terms in the thermodynamic potential.

\section{\label{sec:structure} Crystallographic structure and magnetic vectors.}
\begin{table}
  \centering
  \caption{Ferromagnetic and antiferromagnetic vectors allowed by the symmetry group $D_{2h}^{16}$.} \label{table:LM}
  \begin{tabular}{c}

    $\mathbf{M}=\Spin{1}+\Spin{2}+\Spin{3}+\Spin{4}+\Spin{5}+\Spin{6}+\Spin{7}+\Spin{8}$ \\
    $\vectL{1}=\Spin{1}-\Spin{2}-\Spin{3}+\Spin{4}+\Spin{5}-\Spin{6}-\Spin{7}+\Spin{8}$  \\
    $\vectL{2}=\Spin{1}-\Spin{2}+\Spin{3}-\Spin{4}+\Spin{5}-\Spin{6}+\Spin{7}-\Spin{8}$  \\
    $\vectL{3}=\Spin{1}+\Spin{2}-\Spin{3}-\Spin{4}+\Spin{5}+\Spin{6}-\Spin{7}-\Spin{8}$ \\
    $\vectL{4}=\Spin{1}-\Spin{2}-\Spin{3}+\Spin{4}-\Spin{5}+\Spin{6}+\Spin{7}-\Spin{8}$ \\
    $\vectL{5}=\Spin{1}+\Spin{2}-\Spin{3}-\Spin{4}-\Spin{5}-\Spin{6}+\Spin{7}+\Spin{8}$ \\
    $\vectL{6}=\Spin{1}-\Spin{2}+\Spin{3}-\Spin{4}-\Spin{5}+\Spin{6}-\Spin{7}+\Spin{8}$ \\
    $\vectL{7}=\Spin{1}+\Spin{2}+\Spin{3}+\Spin{4}-\Spin{5}-\Spin{6}-\Spin{7}-\Spin{8}$ \\
  \end{tabular}
\end{table}

\begin{table}
  \centering
  \caption{Magnetic vector components transforming within the same representation.
  The signs in the first column show the effect of the symmetry operations $I$ ($(x,y,z)\rightarrow(-x,-y,-z)$),
   $C^2_z$ ($(x,y,z)\rightarrow(\frac{1}{2}-x,-y,\frac{1}{2}+z)$) and $C^2_y$
   ($(x,y,z)\rightarrow(-x,\frac{1}{2}+y,-z)$), correspondingly.}
  \label{table:sametrans}
  \begin{tabular}{|c|ccc|}
    \hline
    $+++$ & $L_1^x$&$L_2^y$&$L_3^z$ \\
    $++-$ & $L_2^x$&$L_1^y$&$M^z$ \\
    $+-+$ & $L_3^x$&$M^y$&$L_1^z$ \\
    $-++$ & $L_4^x$&$L_6^y$&$L_5^z$ \\
    $--+$ & $L_5^x$&$L_7^y$&$L_4^z$ \\
    $-+-$ & $L_6^x$&$L_4^y$&$L_7^z$ \\
    $+--$ & $M^x$&$L_3^y$&$L_2^z$ \\
    $---$ & $L_7^x$&$L_5^y$&$L_6^z$ \\ \hline
  \end{tabular}
\end{table}

\begin{figure}
  \centering
  \epsfig{file=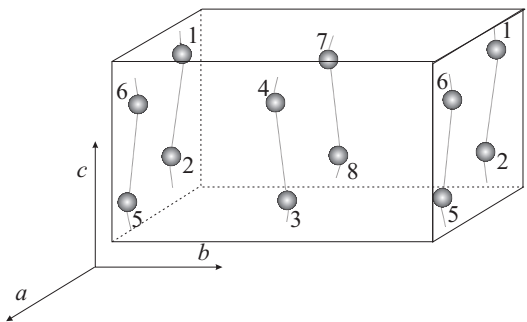, width=0.8\columnwidth, clip=}
  \caption{Position of the magnetic ions in the unit cell.}\label{fig:cell}
\end{figure}

\begin{figure}
  \centering
  \epsfig{file=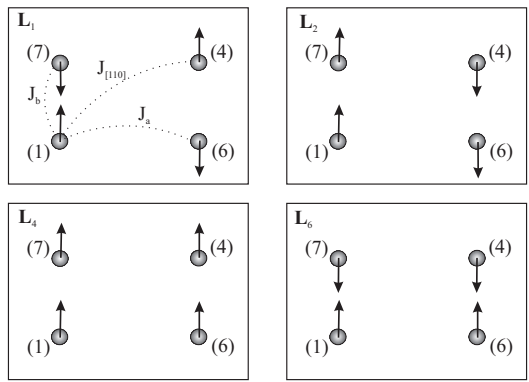, width=0.8\columnwidth, clip=}
  \caption{Relative ordering of spins in the plane perpendicular to the
  chains for different antiferromagnetic vectors.
 The next layer is aligned antiferromagnetically. All spins can be rotated
  simultaneously on an arbitrary angle, including out of plane.}
  \label{fig:l1l2l4l6}
\end{figure}

The compound BaCu$_2$Si$_2$O$_7$ crystallizes in the space group
$Pnma$ ($D_{2h}^{16}$) with four formula units per unit cell i.e.
with 8 Cu$^{2+}$ ions per unit cell. The magnetic ions occupy the
$8d$ positions forming 4 zigzag chains along the $c$ axis (see
Figure \ref{fig:cell}). The lattice parameters are $a=6.862$~\AA,
$b=13.178$~\AA, and $c=6.897$~\AA.

As it follows from neutron scattering, the magnetic unit cell
coincides with the crystallographic one. In the ordered state each
magnetic ion has a non-zero average spin $\mathbf{S}_i$
($i=1\ldots 8$). We will describe the magnetic structure using
linear combinations of these spin
vectors which components transform by the irreducible
representations of the crystal-symmetry group. The space group
$D_{2h}$ exhibits only one-dimensional irreducible
representations. Thus, the  magnetic vector components will remain
the same or change their sign under symmetry operations.

In further analysis we will use the inversion $I$
($(x,y,z)\rightarrow(-x,-y,-z)$) and two sliding axes: $C^2_z$
($(x,y,z)\rightarrow(\frac{1}{2}-x,-y,\frac{1}{2}+z)$) and $C^2_y$
($(x,y,z)\rightarrow(-x,\frac{1}{2}+y,-z)$) as independent
symmetry operations. The fractional cell coordinates of the
Cu$^{2+}$ ion $(1)$ in Figure \ref{fig:cell} are given by
$(\frac{1}{4}-0.028,0.004,\frac{3}{4}+0.044)$ (see
Ref.~\onlinecite{tsukada:si}).

After trivial calculations we obtain 8 magnetic vectors (see Table
\ref{table:LM}). The components of the magnetic vectors that
transform within the same representation are gathered in Table
\ref{table:sametrans}. Here and further on we will use coordinates
with $x\parallel a$, $y\parallel b$ and $z\parallel c$,
respectively. All possible magnetic structures for this symmetry
group (including canted ones) can be analyzed in terms of these
vectors.

A strong in-chain exchange favors ordering of the types
$\vectL{1},\vectL{2},\vectL{4},\vectL{6}$. In these cases
in-chain neighboring spins are antiparallel. We will not consider
other antiferromagnetic vectors in the further analysis. The
relative orientation of the spins is shown for these order types
in Figure \ref{fig:l1l2l4l6}.

If we will neglect the interchain interactions, in classical
($S\gg 1$) approximation the exchange energy for these ordering
types is the same, i.e. the antiferromagnetically ordered state is
four times degenerated. Interchain interactions lift this
degeneration. Taking into account the  exchange-integral values
found by neutron scattering, one can estimate the hierarchy of the
energies of the collinear ordering described by each of these
vectors in the classical approximation.

\begin{equation}\label{eqn:hierarchy}
  \epsilon_6<\epsilon_4<\epsilon_2<\epsilon_1
\end{equation}

Zheludev et al. have found the structure of the magnetic phases by
means of neutron scattering (see
Ref.~\onlinecite{zheludev:struct}). In terms of the above-defined
magnetic vectors the results of this work can be expressed as
follows: At low fields (phase I, $H<H_{c1}$) all spins are aligned
collinearly with antiferromagnetic vector $\vectL{6}\parallel z$.
For the high-field phase (phase III, $H>H_{c2}$) the magnetic
structure was found to be described by the same ordering type
$\vectL{6}$ with the antiferromagnetic vector along the $x$ axis.
The intermediate-field phase (phase II, $H_{c1}<H<H_{c2}$) was
identified as noncollinear antiferromagnetic structure described
by two magnetic vectors $\vectL{6}\parallel y$ and
$\vectL{2}\parallel x$.

The easy-axis magnetic ordering is in agreement with
susceptibility measurements, which show a characteristic kink at
$T_N$ for $\chi_c$. The order of type $\vectL{6}$ is also expected
from classical estimations (see Eqn.~(\ref{eqn:hierarchy})).
However, the identification of the noncollinear phase can be
questioned. First, since nuclear Bragg reflections coincide with
magnetic ones, the determination of magnetic reflection
intensities initially involves a big relative error. As it was
shown in Ref.~\onlinecite{zheludev:struct}, a fit of the
experimental data with the collinear model $\vectL{6}\parallel y$
yields $\chi^2=2.8$, while for the proposed noncollinear
structure $\chi^2=2.4$ was found. Second, from Table
\ref{table:sametrans} one can see that $L_2^x$ and $L_6^y$
transform differently under symmetry operations. This means that
canting towards $L_2^x$ cannot be induced by the internal fields.

\section{Magnetic properties of BaCu$_2$Si$_2$O$_7$.}
The following discussion is organized as follows: in Section
\ref{sec:pd} we will analyze the magnetic phase diagram of
BaCu$_2$Si$_2$O$_7$. The structure of the magnetic phases and the
possible reason of the big values of the relativistic constants
in the thermodynamic-potential expansion will be discussed in
Section \ref{sec:struct}. The behavior of the static
susceptibility above the N\'{e}el point will be treated in Section
\ref{sec:above}.

\subsection{Magnetic phase diagram.\label{sec:pd}}
\begin{figure}
  \centering
  \epsfig{file=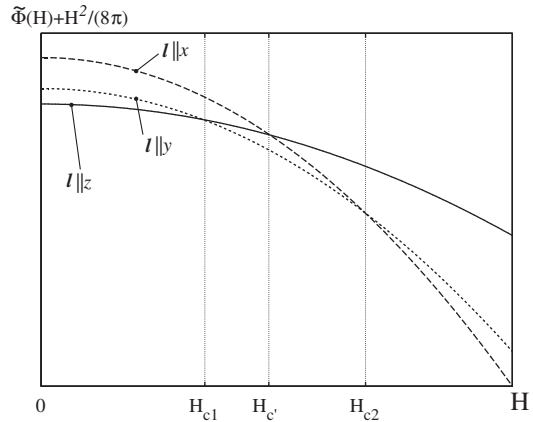, width=0.8\columnwidth, clip=}
  \caption{Field dependences of the thermodynamic potential $\widetilde{\Phi}$ for different
  orientations of the main antiferromagnetic vector. External field $\mathbf{H}\parallel z$.}\label{fig:phi(H)}
\end{figure}

\begin{table}
    \centering
    \caption{Parameters of the thermodynamic potential (see Eqn.~\ref{eqn:exp-oneparam}) at 5K per gramm, calculated from the data of
    Refs.~\onlinecite{tsukada:si},\onlinecite{tsukada:2sf},\onlinecite{poirier:sf3}.}\label{tab:params}
    \begin{tabular}{|c|c|}
    \hline
    $\chi_{\parallel}$&$(1.98\pm 0.08)\cdot10^{-6}$ emu/g\\
    $\chi_{\perp}$&$(3.53\pm 0.03)\cdot10^{-6}$ emu/g\\
    $a_1$&$(845\pm 50)$ erg/g\\
    $a_2$&$(281\pm 13)$ erg/g\\
    $B_1$&$-(1.39\pm 0.07)\cdot10^{-7}$ emu/g\\
    $B_2$&$(1.06\pm 0.14)\cdot10^{-7}$ emu/g\\
    $\xi_x$&$-(1.30\pm 0.8)\cdot10^{-7}$ emu/g\\
    $\xi_z$&$(1.10\pm 0.05)\cdot10^{-6}$ emu/g\\
    \hline
    \end{tabular}
\end{table}

The phase diagram of an antiferromagnet can be analyzed in terms
of the main antiferromagnetic vector (i.e. $\vectL{6}$)
only.\cite{Dzyal} Distortions of the collinear order (i.e.
canting of the sublattices) are of no interest in this subsection,
while we are interested in the phase diagram only. Relativistic
interactions leading to these distortions contribute to the
corresponding relativistic constants, as it will be demonstrated
in the next subsection.

Following the standard procedure, we will write
down an expansion of the thermodynamic potential
$\widetilde{\Phi}(\vectLnorm)$, where  $\vectLnorm=\vectL{6}/|\vectL{6}|$.

The potential $\widetilde{\Phi}$ is defined in such a way
that\cite{landavshic}

\begin{equation}
\frac{\partial\widetilde{\Phi}}{\partial
H}=-\frac{B}{4\pi}=-M-\frac{H}{4\pi} \label{eqn:phidef}
\end{equation}

Since magnetization appears only in the presence of a magnetic
field, the potential expansion can be written via the $\vectLnorm$ and
$\mathbf{H}$ variables.\cite{landavshic} Requiring
invariance of the energy to symmetry operations we obtain:

\begin{eqnarray}
\widetilde{\Phi}(\vectLnorm,\mathbf{H})&=&a_1l_x^2+a_2l_y^2-\frac{\chi_{\parallel}}{2}(\vectLnorm\cdot\mathbf{H})^2-\nonumber\\
&&-\frac{\chi_{\perp}}{2}[\vectLnorm\times\mathbf{H}]^2+C_1(\vectLnorm\cdot\mathbf{H})H_xl_x+\nonumber\\
&&+C_2(\vectLnorm\cdot\mathbf{H})H_yl_y+B_1l_x^2H^2+\nonumber\\
&&+B_2l_y^2H^2-\frac{\xi_z}{2}H_z^2-\frac{\xi_x}{2}H_x^2-\frac{H^2}{8\pi}\label{eqn:exp-oneparam}
\end{eqnarray}

The small parameter in this expansion is the ratio between the
relativistic constants and the exchange constants. The first two
terms ($a_i$) describe the orthorhombic anisotropy, the terms
containing $\chi_{\parallel}$ and $\chi_{\perp}$ describe the
exchange part of the susceptibility --- they do not change under
simultaneous rotation of $\vectLnorm$ and $\mathbf{H}$ on the same
angle. The terms with $\xi_{\alpha}$ describe the relativistic
contribution to the susceptibility which is independent on the
$\vectLnorm$ orientation. Of the higher order terms containing the
magnetic field, we have kept only exchange-relativistic terms
which are a product of exchange part (invariant under simultaneous
rotation of $\vectLnorm$ and $\mathbf{H}$) and relativistic part.
Other higher-order terms are smaller due to the supposed smallness
of the relativistic constants and, thus, can be omitted. Easy-axis
ordering at low fields requires that  $a_1,a_2>0$. Concerning the
anisotropy terms in the expansion (\ref{eqn:exp-oneparam}), it is
also necessary to point out that, if one (or both) of the
anisotropy constants is equal to zero or if $a_1=a_2$, then a
consideration of the higher-order anisotropy terms would become
necessary. However, there are no symmetry reasons, causing these
special values of the anisotropy constants.

Minimization over the orientations of $\vectLnorm$ in the exact
orientations of the external magnetic field demonstrates that the
antiferromagnetic vector $\vectLnorm$ is aligned along one of the
crystallographic axes except maybe for the narrow field interval
near the reorientation transitions where its behavior would be
determined by the higher-order anisotropy terms.

The thermodynamic-potential expressions and magnetic
susceptibilities for different possible orientations of
$\mathbf{H}$ and $\vectLnorm$ are given by:

\begin{itemize}
  \item [$\mathbf{H}\parallel z$] (see Fig.\ref{fig:phi(H)})
  \begin{eqnarray}
  \vectLnorm\parallel z:&\widetilde{\Phi}&=-\frac{\chi_{zz}^{(1)}}{2}H^2-\frac{H^2}{8\pi}\\
  &\chi_{zz}^{(1)}&=\chi_{\parallel}+\xi_z\nonumber\\
  \vectLnorm\parallel y:&\widetilde{\Phi}&=a_2-\frac{\chi_{zz}^{(2)}}{2}H^2-\frac{H^2}{8\pi}\\
  &\chi_{zz}^{(2)}&=\chi_{\perp}-2B_2+\xi_z\nonumber\\
  \vectLnorm\parallel x:&\widetilde{\Phi}&=a_1-\frac{\chi_{zz}^{(3)}}{2}H^2-\frac{H^2}{8\pi}\\
  &\chi_{zz}^{(3)}&=\chi_{\perp}-2B_1+\xi_z\nonumber
  \end{eqnarray}
  \item [$\mathbf{H}\parallel x$]
  \begin{eqnarray}
  \vectLnorm\parallel z:& \widetilde{\Phi}&=-\frac{\chi_{xx}^{(1)}}{2}H^2-\frac{H^2}{8\pi}\\
  &\chi_{xx}^{(1)}&=\chi_{\perp}+\xi_x\nonumber\\
  \vectLnorm\parallel y:& \widetilde{\Phi}&=a_2-\frac{\chi_{xx}^{(2)}}{2}H^2-\frac{H^2}{8\pi}\\
  &\chi_{xx}^{(2)}&=\chi_{\perp}-2B_2+\xi_x\nonumber
  \end{eqnarray}
  \item [$\mathbf{H}\parallel y$]
  \begin{eqnarray}
  \vectLnorm\parallel z:&  \widetilde{\Phi}&=-\frac{\chi_{yy}^{(1)}}{2}H^2-\frac{H^2}{8\pi}\\
  &\chi_{yy}^{(1)}&=\chi_{\perp}\nonumber\\
  \vectLnorm\parallel x:& \widetilde{\Phi}&=a_1-\frac{\chi_{yy}^{(2)}}{2}H^2-\frac{H^2}{8\pi}\\
  &\chi_{yy}^{(2)}&=\chi_{\perp}-2B_1\nonumber
  \end{eqnarray}

\end{itemize}

Now we can find the fields of the spin-reorientation transitions
by equalizing the thermodynamic potentials in the phases with
different orientations of $\vectLnorm$.

\begin{eqnarray}
    \mathbf{H}\parallel z, (\vectLnorm\parallel z \rightarrow \vectLnorm\parallel y)&:&H_{c1}^2=\frac{2a_2}{\chi_{\perp}-\chi_{\parallel}-2B_2}\label{eqn:Hc1}\\
    \mathbf{H}\parallel z, (\vectLnorm\parallel y \rightarrow \vectLnorm\parallel x)&:&H_{c2}^2=\frac{a_1-a_2}{B_2-B_1}\label{eqn:Hc2}\\
    \mathbf{H}\parallel z, (\vectLnorm\parallel z \rightarrow \vectLnorm\parallel x)&:&H_{c'}^2=\frac{2a_1}{\chi_{\perp}-\chi_{\parallel}-2B_1}\label{eqn:H13}\\
    \mathbf{H}\parallel y, (\vectLnorm\parallel z \rightarrow \vectLnorm\parallel x)&:&H_{c3}^2=-\frac{a_1}{B_1}\label{eqn:Hc3}\\
    \mathbf{H}\parallel x, (\vectLnorm\parallel z \rightarrow \vectLnorm\parallel x)&:&H_{c4}^2=-\frac{a_2}{B_2}\label{eqn:Hc4}
\end{eqnarray}

All parameters involved can be estimated using the static
magnetization data from
Refs.~\onlinecite{tsukada:si},\onlinecite{tsukada:2sf} and the
$H_{c3}$ value from Ref.~\onlinecite{poirier:sf3}. Note, however,
that the samples used in the above mentioned papers were
different, thus, some uncertainty is possible. The values of the
parameters calculated from the 5~K data are presented in Table
\ref{tab:params}.
% at 5K
%chi_zz^1=3.085*10^-6 emu/g
%chi_zz^2=4.42*10^-6 emu/g
%chi_zz^3=4.91*10^-6 emu/g
%chi_xx^1=3.40*10^-6 emu/g
%chi_yy^1=3.53*10^-6 emu/g
%error in chi determination 0.05*10^-6 emu/g
%H_c1=20500 Oe H_c2=48000 Oe H_c3=78000 Oe
%error in H determination about 1000 Oe

Since $a_1>a_2$, the hard axis of the magnetization is
established along the $x$ direction. Substituting the found values
to Eqn.~\ref{eqn:H13}, one can ascertain that $H_{c'}>H_{c1}$,
i.e. a two-spin-reorientation transition scenario is more energy
beneficient than a direct transition from $\vectLnorm||z$ to
$\vectLnorm||x$. The positiveness of $B_2$ explains the absence
of a phase transition for $\mathbf{H}\parallel x$: for $a_2>0$ and $B_2>0$
Eqn.~\ref{eqn:Hc4} has no physical solutions.

The anisotropy constants determine the gaps of the
antiferromagnetic resonance spectrum \cite{Kubo}.
Antiferromagnetic resonance on the BaCu$_2$Si$_2$O$_7$ was studied
in Ref.~\onlinecite{Hayn} and two gaps given by $\Delta_1=40$~GHz
and $\Delta_2=76$~GHz were found. The ratio of the AFMR gaps can
be expressed as

\begin{equation}\label{eqn:gapsratio}
  \frac{\Delta_1}{\Delta_2}=\sqrt{\frac{a_2}{a_1}}
\end{equation}

Inserting the values of the anisotropy constants from Table
\ref{tab:params} yields for this ratio $0.57\pm0.04$ which is in
perfect agreement with the value $0.53$ found from the experiment.
The absolute value of the AFMR gap can be approximated
\cite{Kubo} as
\begin{equation}\label{eqn:gapvalue}
  \Delta_1=\gamma\sqrt{2H_{a1}H_e}\sim 30~{\rm GHz}
\end{equation}

which is in reasonable agreement with the experiment, since a good
deal of uncertainty is involved in the estimation of the exchange
field $H_e$ in the case of a low-dimensional magnet. Thus,
expansion (\ref{eqn:exp-oneparam}) correctly describes the
observed spin-reorientation transitions and is in agreement with
other experimental observations.

The additional phase transitions occur as a result of the
competition of anisotropy and relativistic corrections to the
susceptibility. The transition at $H_{c1}$ is a normal spin-flop
transition: all spins are rotated perpendicular to the field
direction and are aligned along the second-easy axis $y$. At this
transition the loss of the easy-axis anisotropy is compensated by
the gain in magnetization energy, since
$\chi_{\perp}>\chi_{\parallel}$. The second phase transition (at
$H_{c2}$) is caused by the competition of the in-plane anisotropy
and difference of the magnetization energy in different
orientations of $\vectLnorm$ due to the relativistic corrections:
The magnetic susceptibility for $\vectLnorm\parallel x$ turns out to be
larger than that for $\vectLnorm\parallel y$, thus, the loss of anisotropy
energy is overcome by the gain in magnetization energy. The
spin-reorientation transition at $\mathbf{H}\parallel y$ is due to the
same effect: relativistic corrections to the susceptibility make
the susceptibility to be the largest for $\vectLnorm\parallel x$,
thus, the gain in magnetization energy overcomes the loss in
anisotropy energy.

Note that the relativistic corrections $B_i$ and $\xi_{\alpha}$
are unusually large in BaCu$_2$Si$_2$O$_7$: $\xi_z$ turns out to
be comparable with $\chi_{\perp}$. The $B_iH^2$ terms normally
should become comparable with the anisotropy constants only at
magnetic fields of the scale of the exchange field. Thus, a
mechanism, which strongly enhances the role of the relativistic
interactions, has to be found. One possibility is the effect of
anisotropic reduction of the spin in different phases. According
to the neutron-scattering data\cite{zheludev:struct}, the average
magnetic moment of the Cu ion changes as $\vectLnorm$ changes its
orientation. Accounting for this effect would result in the same
exchange-relativistic terms in the thermodynamic-potential
expansion. Another possible reason for the large values of these
constants will be discussed in the next section.

It is necessary to recall that the above expansion of the
thermodynamic potential was performed assuming the relativistic
contributions to be small compared to the exchange part. Hence,
higher-order terms have been neglected. However, in principle the
increasing importance of the relativistic interactions should also
lead to an increase of the higher-order terms, which especially
would affect the phase transitions at higher fields ($H_{c2}$,
$H_{c3}$). Moreover, this should give rise to other effects, such
as a nonlinearity of the magnetization, which has not been
reported yet.

We also have to note another possible explanation of the
comparable magnitude of the fields $H_{c1}$ and $H_{c2}$: namely
that, if $a_1=a_2$, then Eqn.\ref{eqn:Hc2} becomes ill-defined. As
it was noted earlier, in this case higher-order (fourth-order)
anisotropy terms have to be considered, and higher-order
anisotropy constants (containing additional relativistic
smallness, compared to $a_{1,2}$) would appear in the numerator
of Eqn.\ref{eqn:Hc2}. The $B_iH^2$ terms should become comparable
with the fourth-order anisotropy terms in fields of the order of
the normal spin-flop field $H_{c1}$. However, we would like to
point out again that there are no symmetry reasons causing the
equivalence of the anisotropy constants, and the values of the
$a_1$ and $a_2$ constants derived from the experimental data are
almost four-fold different. A microscopic consideration of the
interspin interactions, which is beyond of the scope of the
present paper, is necessary to check this probability in detail.
The angular dependence of the transition field $H_{c2}$ also has
to contain information on whether second- or fourth-order
anisotropy terms are responsible for the additional
spin-reorientation transition.

Nevertheless, the present experimental results can be consistently
explained in the framework of our simplified approach. Therefore,
it seems to be reasonable to start again with the assumption of
small relativistic contributions in the following refinement of
the thermodynamic potential with the aim to analyze the magnetic
structure in detail.

\subsection{Structure of the magnetic phases.\label{sec:struct}}

\begin{figure}
  \centering
  \epsfig{file=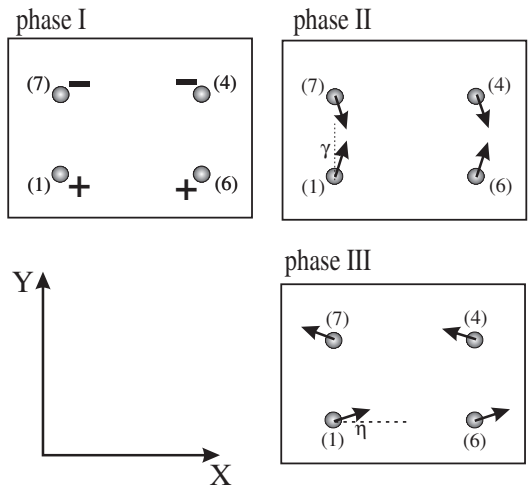, width=0.8\columnwidth, clip=}
  \caption{Proposed structures for all phases (field-induced distortions are omitted):
   phase I: collinear structure  $l_6^z$ (direction of the spins is shown by the sign);
   phase II: weakly noncollinear antiferromagnet $l_6^y$ and $l_4^x$;
   phase III: weakly noncollinear structure $l_6^x$ and $l_4^y$.
  Next layer of spins aligned antiferromagnetically.}\label{fig:noncoll}
\end{figure}

Relativistic interactions, such as the Dzyaloshinskii-Moriya
interaction, can cause small distortions of the collinear
structure, which can be expressed by magnetic vectors. These
distortions, however, have to be compatible with the crystal
symmetry. This means that the components of the magnetic vectors
corresponding to the distortions in the given phase have to
transform in the same way as the corresponding component of the
main magnetic vector, i.e. they have to belong to the same
representation of the symmetry group. The possible types of
canting of the spins can be easily found from Table
\ref{table:sametrans}. We will consider here only distortions
described by the magnetic vectors $\vectL{1}$, $\vectL{2}$ and
$\vectL{4}$, for only these magnetic vectors correspond to the
antiferromagnetic in-chain order. Other distortions would lead to
a violation of the in-chain antiferromagnetic ordering and,
therefore, they are strongly suppressed by the strong in-chain
exchange interaction.

Thus, we obtain
\begin{itemize}
  \item phase I ($\vectL{6}\parallel z$)\\ no canting is allowed by
  symmetry
  \item phase II ($\vectL{6}\parallel y$)\\ $L_4^x$ is allowed
  \item phase III ($\vectL{6}\parallel x$)\\ $L_4^y$ is allowed
\end{itemize}

The structure of the magnetic phases is shown in Figure
\ref{fig:noncoll}. Note that the proposed structure is different
from that of Ref.~\onlinecite{zheludev:struct}, where canting
described by the magnetic vector $\vectL{2}$ was proposed.
However, vector $\vectL{2}$ has components transforming in the
same way as the magnetization, i.e. distortions described by the
components of $\vectL{2}$ can be induced by the magnetic field.

To find the amplitudes of the distortions we have to write an
expansion of the thermodynamic potential including vectors
$\vectL{1}$, $\vectL{2}$ and $\vectL{4}$ and then to minimize
this potential with respect to the components of these vectors.
In this expansion we will normalize all magnetic vectors by the
length of the main magnetic vector $|\vectL{6}|$

\begin{equation}\label{eqn:l-normed}
    \vectLnormI{i}=\vectL{i}/|\vectL{6}|
\end{equation}

Requiring
invariance of the energy to symmetry operations we obtain:

\begin{eqnarray}
\widetilde{\Phi}&=&A_1l_1^2+A_2l_2^2+A_4l_4^2+\beta_1l_6^xl_4^y+\beta_2l_6^yl_4^x+\nonumber\\
&&+\alpha_1l_1^yH^z+\alpha_2l_2^xH^z+\alpha_3l_2^zH^x+\alpha_4l_1^zH^y+\nonumber\\
&&+\frac{\delta\chi}{2}l_4^2H^2-\frac{\chi_{\parallel}'}{2}(\vectLnormI{6}\cdot\mathbf{H})^2-\frac{\chi_{\perp}'}{2}[\vectLnormI{6}\times\mathbf{H}]^2+\nonumber\\
&&+a_1'(l_6^x)^2+a_2'(l_6^y)^2+B_1'(l_6^x)^2H^2+B_2'(l_6^y)^2H^2+\nonumber\\
&&+C_1'(\vectLnormI{6}\cdot\mathbf{H})H^xl_6^x+C_2'(\vectLnormI{6}\cdot\mathbf{H})H^yl_6^y-\nonumber\\
&&-\frac{\xi_z'}{2}H_z^2-\frac{\xi_x'}{2}H_x^2-\frac{H^2}{8\pi}\label{eqn:expansion1}
\end{eqnarray}

We want to emphasize that this expansion is essentially the same
as Eqn.~(\ref{eqn:exp-oneparam}). The only difference is that the
interactions leading to the distortions of the collinear order
are written explicitly, now.

In the expansion given above, the terms containing $A_i$ are of
exchange origin, they describe the loss in exchange energy. All
of $A_i$ are positive. The terms with $\beta_i$ describe the
canting of the antiferromagnetic sublattices (so called weak
antiferromagnetism), those with $\alpha_i$ describe field-induced
distortions. Both $\alpha_i$ and $\beta_i$ parameters arise due to
the microscopic Dzyaloshinskii-Moriya interaction.

Minimization over the components of magnetic vectors will
demonstrate that the amplitudes of $\vectLnormI{1}$,
$\vectLnormI{2}$ and $\vectLnormI{4}$ are proportional to the
ratio of some relativistic constant to the exchange constants
$A_i$. We will suppose here that all relativistic constants are
small compared with exchange constants. Then, the magnitude of
$\vectLnormI{1}$, $\vectLnormI{2}$ and $\vectLnormI{4}$ already
contains relativistic smallness and we can neglect all
higher-order relativistic terms containing these vectors.

Minimization of the (\ref{eqn:expansion1}) yields

\begin{eqnarray}
l_1^y&=&-\frac{\alpha_1}{2A_1}H^z\\
l_1^z&=&-\frac{\alpha_4}{2A_1}H^y\\
l_2^x&=&-\frac{\alpha_2}{2A_2}H^z\\
l_2^z&=&-\frac{\alpha_3}{2A_2}H^x\\
l_4^x&\approx&(-\frac{\beta_2}{2A_4}+\frac{\beta_2}{4A_4^2}\delta\chi\cdot
H^2)l_6^y\label{eqn:l4x}\\
l_4^y&\approx&(-\frac{\beta_1}{2A_4}+\frac{\beta_1}{4A_4^2}\delta\chi\cdot
H^2)l_6^x\label{eqn:l4y}
\end{eqnarray}

The canting angles (see Figure \ref{fig:noncoll}) can be expressed
as follows:

\begin{equation}
\gamma\approx\tan\gamma=l_4^x~~~~~~~~\eta\approx\tan\eta=l_4^y
\end{equation}

Note that the distortions of type $l_2^x$ (which were found in
Ref.~\onlinecite{zheludev:struct}) are possible, but they are
induced by the external field and they are independent on the
orientation of the main antiferromagnetic vector. Thus, they have
to be observed in all phases. Moreover, their amplitude should
increase with increasing magnetic field, while according to the
findings of Ref.~\onlinecite{zheludev:struct} the noncollinearity
is suppressed by the magnetic field in phase II.

On the other hand, the distortions described by the components of
$\vectLnormI{4}$ are consistent with the experimental
observations: they are absent in phase I, and their amplitude in
phases II and III are governed by different constants $\beta_i$.
The fact that the canting in phase III is much smaller than in
phase II forces us to suppose that $\beta_1\ll\beta_2$. A
suppression of the canting in phase II corresponds to
$\delta\chi>0$.

Substituting these results to Eqn.~\ref{eqn:expansion1} we obtain
an equation of the same type as Eqn.~\ref{eqn:exp-oneparam} with
renormalized constants. The comparison of
Eqn.~(\ref{eqn:exp-oneparam}) with the results of the substitution
yields

\begin{eqnarray}
\chi_{\parallel,\perp}&=&\chi_{\parallel,\perp}'+\frac{\alpha_4^2}{2A_1}\label{eqn:contr-begin}\\
\xi_z&=&\xi_z'+\frac{\alpha_1^2}{2A_1}+\frac{\alpha_2^2}{2A_2}-\frac{\alpha_4^2}{2A_1}\\
\xi_x&=&\xi_x'+\frac{\alpha_3^2}{2A_2}-\frac{\alpha_4^2}{2A_1}\\
a_1&=&a_1'-\frac{\beta_1^2}{4A_4}\\
a_2&=&a_2'-\frac{\beta_2^2}{4A_4}\\
B_1&=&B_1'+\frac{1}{8}\left(\frac{\beta_1}{A_4}\right)^2\delta\chi\\
B_2&=&B_2'+\frac{1}{8}\left(\frac{\beta_2}{A_4}\right)^2\delta\chi\label{eqn:contr-end}
\end{eqnarray}

If the relativistic constants are small compared to the exchange
constants (as it was supposed in the expansion
(\ref{eqn:expansion1})), then the distortions caused by
relativistic interactions result in small corrections to the
constants connected with the main antiferromagnetic vector.

In the case of the low-dimensional antiferromagnet, it is another
situation. The antiferromagnetic ordering of the types
$\vectL{1}$, $\vectL{2}$, $\vectL{4}$, and $\vectL{6}$ differs
only by the relative orientation of the spins in the direction
perpendicular to the chain. Thus, the loss in exchange energy for
the canting from one of these order types to the other (i.e.
$A_i$ constants in the expansion (\ref{eqn:expansion1})) is
governed by the weak interchain exchange. As it was mentioned in
the Introduction, in a quasi-one-dimensional magnet the
interchain-exchange interaction can be comparable in strength
with the relativistic interactions. Thus, the relativistic
contributions in Eqns.~\ref{eqn:contr-begin}-\ref{eqn:contr-end}
will not become small.

This argumentation cannot be considered as a strict proof of the
role of low-dimensionality for the large values of the
relativistic constants in Eqn.~\ref{eqn:exp-oneparam}, because
expansion (\ref{eqn:expansion1}) becomes incomplete, if some of
the relativistic constants are comparable to the exchange ones.
However, the fact that in quasi-one-dimensional magnets strong
distortions with small losses of exchange energy are possible, can
result in an increase of the effect due to relativistic
interactions.

\subsection{Susceptibility of BaCu$_2$Si$_2$O$_7$ above the
N\'{e}el point.\label{sec:above}}

The susceptibility of BaCu$_2$Si$_2$O$_7$ above the N\'{e}el point
strongly deviates from a Bonner-Fisher law\cite{BonnerFisher}
characteristic for one-dimensional magnets \cite{tsukada:2sf,
tsukada:si}. It demonstrates a broad maximum near 200~K but then
shows a strong Curie-like increase to lower temperatures. This
increase is anisotropic. It is strongest for $\chi_c$ and smallest
for $\chi_a$.

Such a behavior is typical for antiferromagnets with possible
weak ferromagnetism \cite{borovik:lectures, borovikozhogin}. As it
follows from Table \ref{table:sametrans}, weak ferromagnetism is
allowed for antiferromagnetic order of types $\vectL{1}$ and
$\vectL{2}$
--- these vectors have components transforming in the same way as
the magnetization components.

At the vicinity of the N\'{e}el point the amplitude of the order
parameter is small and we can write down an expansion over the
powers of the order parameter as in the usual Landau theory of
second-order phase transitions:

\begin{eqnarray}
    \widetilde{\Phi}&=&A_1L_1^2+A_2L_2^2+A_4L_4^2+A_6L_6^2+\beta_2L_6^yL_4^x+\nonumber\\
    &&+\beta_1L_6^xL_4^y+\alpha_1L_1^yH^z+\alpha_2L_2^xH^z+\alpha_3L_2^zH^x+\nonumber\\
    &&+\alpha_4L_1^zH^y-\frac{1}{2}\chi_pH^2-\frac{H^2}{8\pi}\label{eqn:expansion2}
\end{eqnarray}

Here the $A_i$ terms are of exchange origin, those with $\alpha$
and $\beta$ are responsible for the canting, $\chi_p$ is the
paramagnetic-state susceptibility. We neglect anisotropy terms
and higher-order terms. Here we have used designations for the
terms similar to those in Eqn.~\ref{eqn:expansion1}. Note,
however, that coefficients with the same designations are
different in Eqn.~\ref{eqn:expansion1} and
Eqn.~\ref{eqn:expansion2}.

At the transition point the factors $A_i$ change their sign: it is
positive above the transition temperature and negative below. In
the vicinity of the corresponding N\'{e}el temperatures

\begin{equation}
A_i=\lambda_i(T-T_N^{(i)}),~~~~~\lambda>0\label{eqn:Ai(T)}
\end{equation}

At the N\'{e}el point order of type $\vectL{6}$ is established. We
will suppose that in some temperature range above $T_N$ the
values of $A_1$ and $A_2$ continue to increase with increasing
temperature and, thus, can be approximated as

\begin{equation}\label{eqn:Ai-approx}
  A_{1,2}=A_{1,2}^{(0)}+\lambda_{1,2}'(T-T_N)~~~~~~~~A_{1,2}^{(0)},\lambda_{1,2}'>0
\end{equation}

To find the values of the magnetic vector components we have to
minimize expansion (\ref{eqn:expansion2}). Here we will restrict
our analysis to the case of $T>T_N$. Above the N\'{e}el point we have
$A_i>0$, and in absence of an external field $\mathbf{H}$
the minimum of the potential (\ref{eqn:expansion2}) corresponds to
$|\vectL{i}|=0$. If a magnetic field is applied, weak
antiferromagnetic order parameters of the types $\vectL{1}$ or $\vectL{2}$
are induced. All other magnetic vectors are zero.

The existence of field-induced order parameters leads to an
additional contribution to the susceptibility. Using
Eqn.~\ref{eqn:Ai-approx}, we obtain for the magnetic
susceptibilities:

\begin{eqnarray}
    \chi_{xx}&=&\chi_p+\frac{\alpha_3^2}{2(A_{2}^{(0)}+\lambda_2'(T-T_N))}\label{eqn:chipmxx1}\\
    \chi_{yy}&=&\chi_p+\frac{\alpha_4^2}{2(A_{1}^{(0)}+\lambda_1'(T-T_N))}\label{eqn:chipmyy1}\\
    \chi_{zz}&=&\chi_p+\frac{\alpha_1^2}{2(A_{1}^{(0)}+\lambda_1'(T-T_N))}+\nonumber\\
    &&~~~~~~~~~~~~~~~+\frac{\alpha_2^2}{2(A_{2}^{(0)}+\lambda_2'(T-T_N))}\label{eqn:chipmzz1}
\end{eqnarray}

I.e. with decreasing temperature the susceptibilities should
increase. The amplitudes of this increase in different
orientations \cite{tsukada:si} allow to conclude that

\begin{equation}
\alpha_1^2+\alpha_2^2>\alpha_4^2\gg\alpha_3^2
\end{equation}

The strongest effect for $\chi_{zz}$ is in accordance with the
large relativistic contribution $\xi_z$ in
Eqn.~\ref{eqn:exp-oneparam}.

\section{Conclusions}

Starting from a symmetry approach we have analyzed the magnetic
phase diagram and the corresponding magnetic structures of the
antiferromagnet BaCu$_2$Si$_2$O$_7$. In the first step, we have
obtained a thermodynamic potential in terms of the main
antiferromagnetic vector only, demonstrating additional
spin-reorientation transitions, which are due to the unusually
strong relativistic terms in the expansion of the thermodynamic
potential. The strength of the relativistic terms is probably
connected to the low-dimensionality of the compound.

In the second step, we have refined the thermodynamic potential by
including the magnetic vectors, which describe distortions of the
magnetic structure compatible with the crystal symmetry. Based on
our analysis, we proposed a structure of the magnetic phases (cf.
Fig. \ref{fig:noncoll}), which is different from that in
Ref.~\onlinecite{zheludev:struct}.

In addition, we were able to explain the deviations of the
susceptibility from the Bonner-Fisher law on approaching the
3D-magnetic order. The analysis of the susceptibility behavior
above $T_N$ also shows the importance of the relativistic
interactions for the understanding of the properties of
BaCu$_2$Si$_2$O$_7$.

The strong enhancement of the relativistic effects should result
in other interesting properties of this compound, such as
nonlinear magnetization. Thus, further experimental and
theoretical studies are necessary to achieve a deeper insight
into the unusual properties of BaCu$_2$Si$_2$O$_7$.

\acknowledgements

This work was supported by the joint grant of the Russian
Foundation for Basic Research and Deutsche Forschungsgemeinschaft
(DFG) No.01-02-04007. The authors thank M.Zhitomirsky, A.I.Smirnov
and S.S.Sosin for valuable discussions.

\end{document}